# Text Classification Using Hybrid Machine Learning Algorithms on Big Data

[1]Asogwa D.C, [2]Anigbogu S.O, [3]Onyenwe I. E and [4]Sani F.A,
[1,2,3]Department of Computer Science, Faculty of Physical Sciences, Nnamdi Azikiwe University, Awka, Anambra State, Nigeria
[4]Federal Polytechnic Idah, Kogi state, Nigeria

*Abstract:* Recently, there are unprecedented data growth originating from different online platforms which contribute to big data in terms of volume, velocity, variety and veracity (4Vs). Given this nature of big data which is unstructured, performing analytics to extract meaningful information is currently a great challenge to big data analytics. Collecting and analyzing unstructured textual data allows decision makers to study the escalation of comments/posts on our social media platforms. Hence, there is need for automatic big data analysis to overcome the noise and the non-reliability of these unstructured dataset from the digital media platforms. However, current machine learning algorithms used are performance driven focusing on the classification/prediction accuracy based on known properties learned from the training samples. With the learning task in a large dataset, most machine learning models are known to require high computational cost which eventually leads to computational complexity. In this work, two supervised machine learning algorithms are combined with text mining techniques to produce a hybrid model which consists of Naïve Bayes and support vector machines (SVM). This is to increase the efficiency and accuracy of the results obtained and also to reduce the computational cost and complexity. The system also provides an open platform where a group of persons with a common interest can share their comments/messages and these comments classified automatically as legal or illegal. This improves the quality of conversation among users. The hybrid model was developed using WEKA tools and Java programming language. The result shows that the hybrid model gave 96.76% accuracy as against the 61.45% and 69.21% of the Naïve Bayes and SVM models respectively.

## I. INTRODUCTION

Online social networks (OSN) and micro-blogging websites are attracting internet users more than any other kind of websites. Their contents are rapidly growing in speed, volume, velocity and variety, constituting a very interesting example of big data. Big data have been attracting the attention of researchers who are interested in the analysis of people's opinions and the structure/distribution of users in digital media platforms. These websites offer an open space for people to discuss and share thoughts and opinions and as such, the nature and the huge number of posts, comments and messages exchanged make it almost impossible to control their content. Also given the different backgrounds, cultures and believes, many people tend to use illegal comments/messages when discussing with people who do not share the same backgrounds. Illegal comment/message is a speech that is intended to insult, offend or intimidate a person because of some trait (as race, religion, sexual orientation, national origin and so on). Collecting and analyzing these data allows decision makers to study the escalation of such comments/messages. Hence, there is need for automatic big data analysis to overcome the noise and the non-reliability of these unstructured dataset from the digital media platforms.There are different existing models for machine learning on big data and they include: Decision Tree based model, Linear regression based model, Neural Network, Bayesian Network, Support vector machines, Nearest Neighbor, KNN, Naïve Bayes and so on. To reduce the individual limitations of basic machine learning models, a hybrid model is proposed in this work. This will also help to exploit their different generalization mechanisms and improve the expected output of semi structured sequential data. The strengths of Naïve Bayes are that it is very simple and highly scalable on large dataset and can learn incrementally because it counts the observed variables and update the probability distribution table.It has also proven to be one of the best in text classification, spam filtering, hybrid recommender system, online application and simple emotion modeling ( Ravindra Bachate, 2016). Super vector machines (SVM) are also ranked as one of the best off the shelf supervised learning algorithm as they provide superior generalization performance, requires less examples for training and can tackle high dimensional data with the help of kernels (Yinglie Tian et al,2012). SVM also tends to perform well for classification of text because of itsability to generalize into high dimensions, which is often the case with text categorization.

Based on these, two supervised machine learning algorithms, SVM and Naïve Bayes are combined in this work to classify the comments in big data emanating from different blogs.

## II. RELATED WORKS

Junfei et al (2016) presented a survey of the latest advances in researches on machine learning for big data processing; they reviewed machine learning techniques and highlighted some promising learning methods such as representation learning, deep learning, distributed and parallel learning, transfer learning, active learning, and kernel-based learning. They concluded by investigating the close connections of machine learning with signal processing techniques for big data processing.

Moorthy& Gandhi, (2018)reviewed security issues in big data and evaluated the performance of Machine Learning (ML) and Deep Learning (DL) algorithm in a critical environment. They concluded that both ML and DL have security issues.

Perry, (2013) proposed the selection of appropriate machine learning methodologies that can offer substantial improvements in accuracy and performance. He proffered that even at this early stage in testing machine learning on conflict prediction, full models offer more predictive power than simply using a prior outbreak of violence as the leading indicator of current violence.

This suggests that a refined data selection methodology combined with strategic use of machine learning algorithms could indeed offer a significant addition to the early warning toolkit. His work only provided a feasibility study on conflict prediction

Asha et al,. (2013), stated that machine learning algorithms (MLA) are sequential and recursive and the accuracy of MLA's rely on size of the data (i.e., greater the data the more accurate is the result). Absence of a reliable framework for MLA to work for big data has made these algorithms to cripple their ability to reach the fullest potential. Hadoop is one such framework that offers distributed storage and parallel data processing. The Existing problem to implement MLA on Hadoop is that the MLA's need data to be stored in single place because of its recursive nature, but Hadoop does not support data sharing. They proposed an approach to build Machine Learning models for recursive MLA's on Hadoop so that the power of Machine Learning and Hadoop can be made available to process big data, and then compared the performance of ID3 decision tree algorithm, K-means clustering algorithm and K-Nearest Neighbor algorithm on both serial implementation and parallel implementation using Hadoop

Awad (2012), used machine learning algorithms like SVM, KNN and GIS to perform a behavior comparison on web pages classifications problem. From their experiment, they concluded that when using SVM with small number of negative documents to build the centroids it has the smallest storage requirement and the least on line test computation cost. But almost all GIS with different number of nearest neighbors have an even higher storage requirement and on line test computation cost than KNN.

Bart (2017), noted that financial institutions (FIs) are looking to more powerful analytical approaches in order to manage and mine increasing amounts of regulatory reporting data and unstructured data, for purposes of compliance and risk management. He showed that the ability of machine learning methods to analyze very large amounts of data, while offering a high granularity and depth of predictive analysis, can improve analytical capabilities across risk management and compliance areas in financial institutions

Apoorva & Anupam (2017), said that machine learning have been playing a pivotal role for decades in almost every field like Finance, Weather forecasting, Medical science, Engineering optimization problems, Software cost prediction, Pattern matching etc-. With the improvement in hardware technology and availability of huge amount of training data, deep learning techniques have emerged as the most important machine learning algorithms. Some of the popular deep learning techniques are CNN, RNN, LSTM, etc-. Although deep learning techniques give far better results than the shallow traditional machine learning algorithms, there are some issues associated with them like: the need for huge amount of data for training, need for expensive hardware resources and the requirement for very high running time. They concluded that despite these limitations, deep learning algorithms are very useful and are being widely used for solving various problems related to different fields.

LinaZhouet al (2017) introduced a framework of ML on big data (MLBiD) to guide the discussion of its opportunities and challenges. The framework is centered on ML which follows the phases of preprocessing, learning, and evaluation. In addition, the framework is also comprised of four other components, namely big data, user, domain, and system. The phases of ML and the components of MLBiD provide directions for identification of associated opportunities and challenges and open up future work in many unexplored or under explored research areas.

Mohammed et al, (2016) proposed an intuitive but yet simple machine learning (ML) approach that consists of two generic algorithms augmenting one another to solve problems they are not designed to solve. They attempted to augment the architecture of traditional Artificial Neural Network (ANN) with a state machine acting as a form of short term memory in addition to help divide the work amongst multiple modular ANNs through transitioning from state to state.

Alexyet al (2016) presented a comprehensive review of the most effective content-based e-mail spam filtering techniques. They focused primarily on Machine Learning-based spam filters and their variants, and reported on a broad review ranging from surveying the relevant ideas, efforts, effectiveness, and the current progress. They concluded by measuring the impact of Machine Learning-based filters and explore the promising offshoots of latest developments.

Yenala, et al (2017) proposed a novel deep learning-based technique for automatically identifying inappropriate languages in a text. They focused on solving the problem in two application scenarios: query completion suggestions in search engines and users conversations in messengers. They proposed a novel deep learning architecture called convolutional bi-directional LSTM (C-BiLSTM) which combines the strengths of both convolution neural network (CNN) and Bi-directional LSTMs (BLSTM).

Vandersmissen, et al (2012) reported that a combination of support vector machine (SVM) and word list-based classifier performs best and observed their models perform badly with less context which is usually the case with search queries as well. However, they specifically focused on the detection of sexually inappropriate and racist messages only.

Georgioset al (2018) addressed the important problem of discerning hateful content in social media. They proposed a detection scheme that is an ensemble of recurrent neural network (RNN) classifiers and it incorporates various features associated with user related information such as the users tendency towards racism or sexism. The scheme can successfully distinguish racism and sexism messages from normal text and achieve higher classification quality.

Arthur (2018) discussed about the implementation of a simple anti-spam control based on the famous naïve Bayesian classification algorithm that can be actively used to locate and filter out those texts (E-mails, SMS, …) from a local messages database that most likely might contain spam or other unsolicited data.

Davidson T. et al, 2017 automatically classified tweets into hate speech, offensive and neither of the above. They compared among logistic regression, naïve Bayes, decision trees, random forest and SVM. They concluded that logistic regression gave the best prediction. They never combined any of the algorithms.

Hajime Watanabe et al, 2018, proposed a new method to detect hate speech in twitter and classified the tweets into hateful, offensive and clean. They used only one algorithm J48graft.

Georgios K. Pitsilis et al, 2018 used deep learning algorithm to address the important problem discerning hateful contents in social media. They were able to distinguish racism and sexism messages from normal text.

Grondahl et al, 2018 reproduced seven state-of-the-art hate speech detection models from prior work, and showed that they perform well only when tested on the same type of data they were trained on. Based on these results, they argued that for successful hate speech detection, model architecture is less important than the type of data and labeling criteria. They further showed that all proposed detection techniques are brittle against adversaries who can (automatically) insert typos, change word boundaries or add in-nocuous words to the original hate speech.

Durairaj M. et al, 2017 combined K nearest neighbor, Support vector machine (SVM) and Naïve Bayes with text mining techniques to classify some text document. This system performed very well with the accuracy of 83.5% but still had some challenges. These challenges include addressing the problems like handling large text corpora, similarity of words in text documents and association of text documents with a subset of class categories.

### III. METHODOLOGY

The methodology adopted in this work is the Object-Oriented Analysis and Design (OOADM) with text mining techniques. The hybrid model was developed using WEKA tools which is a collection of machine learning algorithms for data mining tasks. It contains tools for data pre-processing and classification which makes it very suitable for the development of the new hybrid model. The codes were developed in JAVA programming language and a confusion matrix used to evaluate the accuracy

*A. System Design and Implementation*

Text classification involves preprocessing of the unstructured text documents retrieved from the different web blogs used. After the preprocessing comes the feature extraction of labeled corpus (machine learning classifier for training data) followed by the model selection and classifier. These steps can be seen in figure 4.1

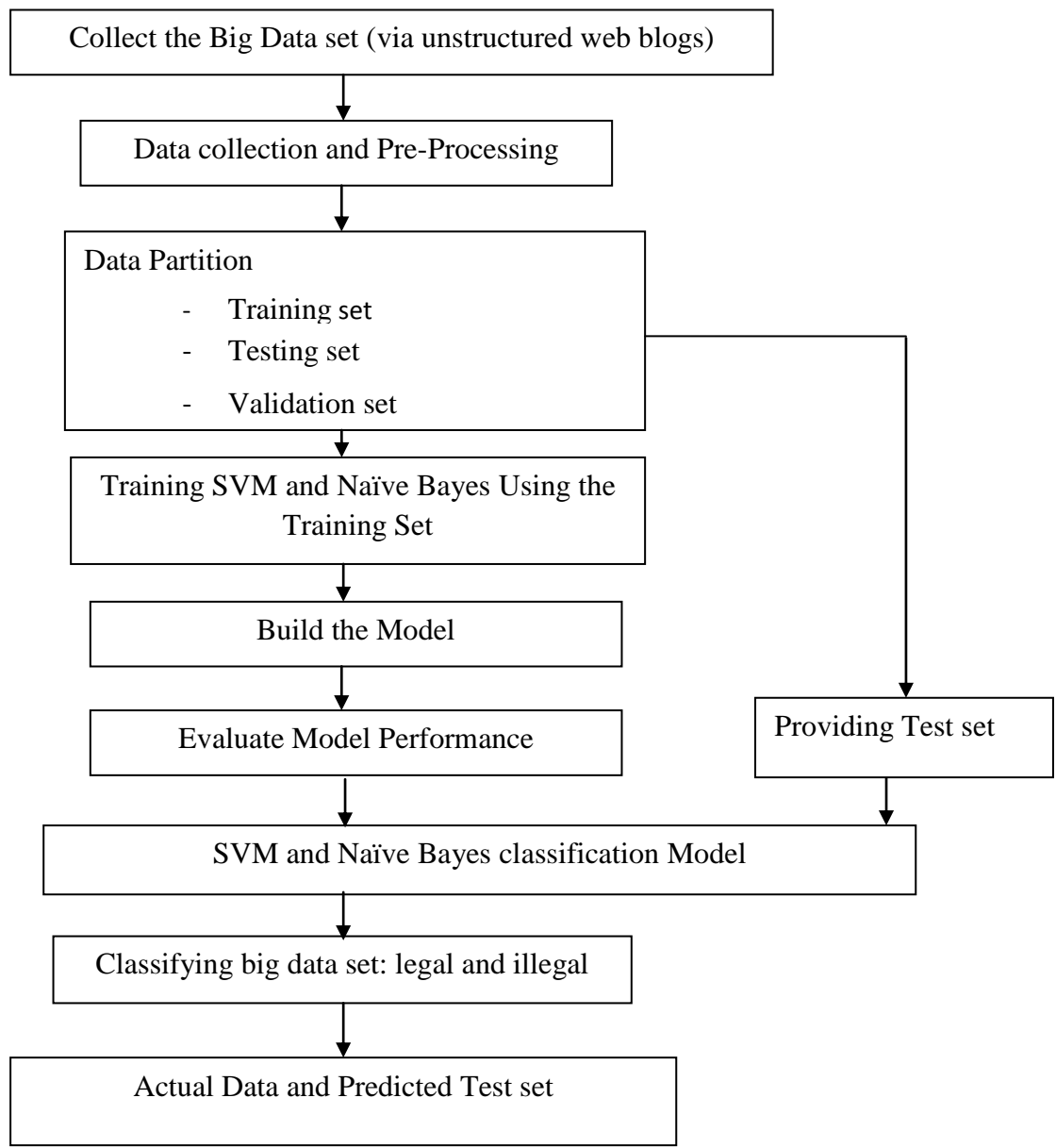

Figure 3.1 Data Flowdiagram of the proposed system

### i) Data set collection

The dataset was collected from the online social web blogs. These blogs are where people with different backgrounds, culture, beliefs and mind sets read trending news and make comments at any point in time. These comments/messages were retrieved as the unstructured data set. The unstructured data set were collected from the following links which form the basis of the big data analytics:

a) https://foreignpolicy.com/comments_view/?view_post_comments=https://foreignpolicy.com/2016/06/14/if-islam-is-a-religion-of-violence-so-is-christianity/
b) http://theconversation.com/challenging-the-notion-that-religion-fosters-violence-85677
c) https://www.nieuwwij.nl/english/karen-armstrong-nothing-islam-violent-christianity/
d) https://www.kaggle.com/c/detecting-insults-in-social-commentary
e) http://www.religionforums.org/ .

These web-blogs allow anonymous user comments on articles. Their policy on which comments are allowed is not very restrictive, meaning a lot of abusive and inappropriate comments are still available online. This fact is of great importance in order to havea representative data set.These comments are unlabeled and the training data must first be manually labeled to train a machine learning algorithm to detect legal/illegal comments.

### ii) Preprocessing

The preprocessing stage is a major task of cleaning up the data gathered from unstructured web blog, by labeling those data into their respective classes. This may involve 85% of human effort to clean up the dataset in order to be used for building the model and reduce the error during the binary classification

### iii) Feature set Extraction

Feature extraction is a dimensionality reduction process where the dataset is reduced to more manageable features for processing while still accurately and completely describing the original dataset. It is intended to be informative and non-redundant, facilitating the subsequent learning and generalization steps in some cases leading to better human interpretations. Following are different feature selection model while dealing with big data classification.

## IV. RESULTS AND DISCUSSIONS

**Naive Bayes Classifier**

Model Information

=================

```
Correctly Classified Instances       1499           62.3544 %
Incorrectly Classified Instances      905           37.6456 %
Kappa statistic                       0.2691
K&B Relative Info Score              53356.4999 %
K&B Information Score                 525.3437 bits     0.2185 bits/instance
Class complexity | order 0           2366.6915 bits     0.9845 bits/instance
Class complexity | scheme           16713.8529 bits     6.9525 bits/instance
Complexity improvement     (Sf)    -14347.1614 bits    -5.968  bits/instance
Mean absolute error                   0.3772
Root mean squared error               0.6015
Relative absolute error              77.0996 %
Root relative squared error         121.6024 %
Total Number of Instances            2404
```

Table 4.9: Naïve Bayes Algorithm Detailed Accuracy by Class

|  | TP Rate | FP Rate | Precision | Recall | F-Measure | MCC | ROC Area | PRC Area | Class |
|---|---|---|---|---|---|---|---|---|---|
|  | 0.518 | 0.235 | 0.748 | 0.518 | 0.612 | 0.286 | 0.683 | 0.757 | Illegal Messages |
|  | 0.765 | 0.482 | 0.542 | 0.765 | 0.634 | 0.286 | 0.683 | 0.559 | Legal Messages |
| Weighted Avg | 0.624 | 0.340 | 0.660 | 0.624 | 0.622 | 0.286 | 0.683 | 0.672 |  |

Table 4.10 Confusion Matrix/Contingency Table

| A | B | <-- classified as | |
|---|---|---|---|
| **714** | 664 | a\| | **IllegalMessages** |
| 241 | **785** | b\| | **legalMessages** |

Table 4.9 and 4.10 shows the general interpretation of results with the confusion matrix for performance evaluation. The evaluation of results for Naïve Bayes as shown in table 4.9 is shown below to confirm the results produced.

Accuracy is measured by the area under the receiver operating characteristics (ROC) curve. An area of 1 under the ROC represents a perfect test; a worthless test is denoted by an area of 0.5. The area under ROC measures the capability of the test to correctly classify the signals with the big data analytics and the normal signals as shown in figure 4.9 below.

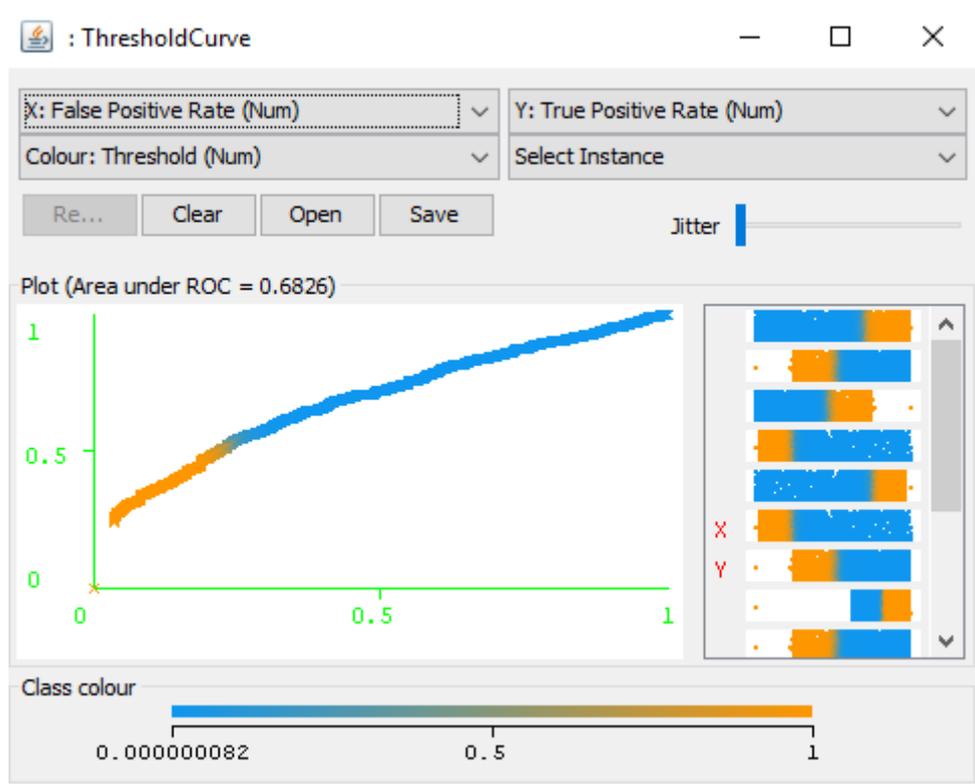

Figure 4.9: Receiver Operating Curve

**Support Vector Machine Classifier**

Model Information

==================

```
Correctly Classified Instances     2004         83.3611 %
Incorrectly Classified Instances    400         16.6389 %
Kappa statistic                     0.6628
K&B Relative Info Score             157853.634  %
K&B Information Score               1554.2139 bits     0.6465 bits/instance
Class complexity | order 0          2366.6915 bits     0.9845 bits/instance
Class complexity | scheme           429600    bits   178.7022 bits/instance
Complexity improvement    (Sf)     -427233.3085 bits  -177.7177 bits/instance
Mean absolute error                 0.1664
Root mean squared error             0.4079
Relative absolute error             34.0062 %
Root relative squared error         82.4705 %
Total Number of Instances           2404
```

Table 4.11a: SVM Detailed Accuracy by Class

|  | TP Rate | FP Rate | Precision | Recall | F-Measure | MCC | ROC Area | PRC Area | Class |
|---|---|---|---|---|---|---|---|---|---|
|  | 0.830 | 0.162 | 0.873 | 0.830 | 0.851 | 0.664 | 0.834 | 0.822 | Illegal Messages |
|  | 0.838 | 0.170 | 0.786 | 0.838 | 0.811 | 0.664 | 0.834 | 0.728 | Legal Messages |
| Weighted Avg | 0.834 | 0.165 | 0.836 | 0.834 | 0.834 | 0.664 | 0.834 | 0.782 |  |

Table 4.12 Confusion Matrix/Contingency Table

| A | B | <-- classified as | |
|---|---|---|---|
| **1144** | 234 | a| | IllegalMessages |
| 166 | **860** | b| | legalMessages |

Table 4.11 and 4.12a above shows the general interpretation of results with the confusion matrix for SVM evaluation of results on big data analytics. The evaluation of results for SVM as shown in table 4.11 is shown below to confirm the results produced.

**Results Evaluation of Hybrid Machine Learning Model for Big data Analytics and Discussion**

Correctly Classified Instances        508            96.7619 %
Incorrectly Classified Instances      17             3.2381 %
Kappa statistic                       0.864
Mean absolute error                   0.0318
Root mean squared error               0.1619
Relative absolute error               13.0954 %
Root relative squared error           46.5217 %
Total Number of Instances             525

Table 4.11b: Hybrid model Detailed Accuracy by Class

|  | TP Rate | FP Rate | Precision | Recall | F-Measure | MCC | ROC Area | PRC Area | Class |
|---|---|---|---|---|---|---|---|---|---|
|  | 0.865 | 0.016 | 0.901 | 0.865 | 0.883 | 0.864 | 0.994 | 0.970 | legalmessages |
|  | 0.984 | 0.135 | 0.978 | 0.984 | 0.981 | 0.864 | 0.994 | 0.999 | illegalmessage |
| Weighted Avg | 0.968 | 0.118 | 0.118 | 0.967 | 0.968 | 0.967 | 0.864 | 0.994 | 0.995 |

| A | B | <-- classified as | |
|---|---|---|---|
| **64** | 10 | a| | legalMessages |
| 7 | **444** | b| | illegalmessage |

**Accuracy rates**

The accuracies of two classifiers have been represented in Table 4.13. Thus, training data set "SVM" has the best accuracy rate of 69.21% and training data set "Naïve Bayes" has an accuracy rate of 61.43% which is only next to that of "SVM". Also on the test set, "SVM" has the best accuracy rate of 83.3611% and test data set "Naïve Bayes has 62.3544%. Based on these results, the accuracy of hybridization of SVM + Naïve for the big data analytics can be seen in the table below.

Table 4.13 Accuracy by classifier/algorithm

| Model Accuracy by class for Naïve Bayes and Support Vector Machine | Performance Rate |
|---|---|
| *Training Set* | Accuracy Rate |
| SVM | 69.21 |
| Naïve Bayes | 61.43 |
|  |  |

| Test Set | Accuracy Rate |
|---|---|
| SVM | 83.3611 |
| Naïve Bayes | 62.3544 |
|  |  |
| *Training Set/Test Set on Binary Classification* | Accuracy Rate |
| SVM + Naïve Bayes | 96.7619 |

**CONCLUSION**

The propagation of illegal messages on social media has been increasing significantly in recent years. This may be as a result of the anonymity and mobility of such platforms, as well as the changing political climate from many places in the world. Despite substantial effort from law enforcement departments, legislative bodies as well as millions of investment from social media companies, it is widely recognized that effective counter measures rely on automated semantic analysis of such content.

A crucial task in this direction is the detection and classification of such messages/comments based on its targeting characteristics.This work makes several contributions to state of the art in this research area. A thorough data analysis was carried out to understand the extremely unbalanced nature and the lack of discriminative features of illegal content in the unstructured dataset one has to deal with in such tasks. However, it is always difficult to clearly decide on a sentence whether it contains illegal words or not if the message is hiding behind sarcasm or if no clear words showing illegal, racism or stereotyping exist. Furthermore, online social networks are full of ironic and joking content that might seem illegal which in reality is not.

A new hybrid model combining the Naïve Bayes and super vector machines is used for more accuracy and efficiency. The methods were thoroughly evaluated on a very large collection of web blog datasets for illegal messages to show that they can be particularly effective on detecting and classifying the contents. The results obtained from the different algorithms really shows that each of them can perform the analysis very well but the combination of the two algorithms actually yielded much more accuracy and efficiency. Clearly, combining models improves model accuracyand reduces model variance, and the more models combined the better the result.. However, determining which individual models combine best from training results only is difficult—there is no clear trend. Simply selecting the best individual models does not necessarily lead to a better combined result. Combining models across algorithm families reduces error compared to the best single models.

*A. Suggestion for Further Research*

The system can be improved and advanced based on the following:

1. To explore other methods or algorithms that aim at compensating the lack of training data in supervised learning tasks. Methods such as transfer learning could be potentially promising as they study the problem of adapting supervised models trained in a resource-rich context to a resource-scarce context.
2. To investigate whether features discovered from one illegal class can be transferred to another thus enhancing the training of each other.
3. To train more models and combine their predictions for more robust and accurate models.